\documentclass[onecolumn]{article}
\usepackage{amssymb}
\usepackage[margin=1in]{geometry}
\usepackage{graphicx}
\usepackage{epstopdf}
\usepackage{color}

\usepackage{lineno}

\begin{document}
\begin{center}
	\Large Direct identification of dilute surface spins on Al$_2$O$_3$: Origin of flux noise in quantum circuits\\\vspace{2mm}\normalsize
	S. E. de Graaf$^{1\dagger}$, A. A. Adamyan$^2$, T. Lindstr\"om$^1$, D. Erts$^3$, S. E. Kubatkin$^2$, A. Ya. Tzalenchuk$^{1,4}$, and A. V. Danilov$^2$\\\vspace{2mm}\footnotesize
	$^1$ National physical laboratory, Hampton Road, Teddington, TW11 0LW, UK\\
	$^2$ Department of Microtechnology and Nanoscience, Chalmers University of Technology, SE-412 96 G\"oteborg, Sweden\\
	$^3$ Institute of Chemical Physics, University of Lativa, LV 1586, Latvia\\
	$^4$ Royal Holloway, University of London, Egham, TW20 0EX, UK\\
	$^\dagger$ sdg@npl.co.uk
	\end{center}

{\bf It is universally accepted that noise and decoherence affecting the performance of superconducting quantum circuits are consistent with the presence of spurious two-level systems (TLS) (see \cite{paladino2014} for a recent review). In recent years bulk defects have been generally ruled out as the dominant source, and the search has focused on surfaces and interfaces \cite{wang2015b, gao2008}. Despite a wide range of theoretical models \cite{koch2007,holder2013,faoro2015,lee2014, wu2012,faoro2008,wang2015,gordon2014,faoro2012} and experimental efforts \cite{wang2015b,sendelbach2008, burnett2014, anton2013}, the origin of these surface TLS still remains largely unknown, making further mitigation of TLS induced decoherence extremely challenging.

Here we use a recently developed \cite{aplpaper} on-chip electron spin resonance (ESR) technique that allows us to detect spins with a very low surface coverage. We combine this technique with various surface treatments specifically to reveal the nature of native surface spins on Al$_2$O$_3$ -- the mainstay of almost all solid state quantum devices \cite{paladino2014}. On a large number of samples we resolve three ESR peaks with the measured total paramagnetic spin density $n=2.2\times 10^{17}$m$^{-2}$, which matches the density inferred from the flux noise in SQUIDs \cite{koch2007,sendelbach2008}. We show that two of these peaks originate from physisorbed atomic hydrogen which appears on the surface as a by-product of water dissociation \cite{lu2015}. We suggest that the third peak is due to molecular oxygen on the Al$_2$O$_3$ surface captured at strong Lewis base defect sites \cite{bedilo2014,medvedev2012}, producing charged O$_2^-$. 
These results provide important information towards the origin of charge and flux noise in quantum circuits. Our findings open up a whole new approach to identification and controlled reduction of paramagnetic sources of noise in solid state quantum devices.
}

Due to its many unique properties Al$_2$O$_3$ is important for a wide range of emerging technologies. Specifically for superconducting devices single-crystal Al$_2$O$_3$ is the best low-loss substrate for high quality planar resonators and is present as an amorphous dielectric oxide layer in Josephson junctions made from Al. 
Here we explicitly focus on Al$_2$O$_3$ in the context of quantum computing technologies and the presence of unwanted material defects. A variety of models have been suggested that could explain the ubiquitous 1/f noise and high level of decoherence found in solid state quantum devices \cite{paladino2014}. Noise in superconducting resonators and charge qubits is typically derived from a bath of electric dipoles coupling to the device, while in Superconducting Quantum Interference Devices (SQUIDs) and flux qubits a bath of magnetic dipoles results in similar  1/f noise and decoherence.
To explain the observed flux noise in SQUIDs it has been suggested that surface functionalisation results in electron spin exchange via hyperfine interactions \cite{wu2012} or a bath of paramagnetic ions \cite{lee2014}. Recent experiments  point towards the formation of a surface spin glass and RKKY interactions \cite{faoro2008,sendelbach2008} and spin diffusion \cite{anton2013,lanting2014}.
Widely different types of spin defects has been suggested as possible sources of flux noise, for example surface dangling bonds \cite{sendelbach2008,desousa2007}, adsorbed molecules \cite{lee2014, wang2015, kumar2016} or intrinsic nuclear spins \cite{wu2012} and recent results provide strong experimental evidence for paramagnetic species being responsible for flux noise \cite{martinisfluxnoise2016}.
Similarly, a wide range of models seek to explain the electrically coupled TLSs in resonators and charge qubits where the latest results indicate similar mechanisms at interfaces \cite{wang2015b,gao2008,faoro2015,faoro2012, burnett2014}. Ab inito studies have suggested several possible candidates for charge coupled TLS, such as hydrogen and hydroxyl defects \cite{holder2013}, or tunnelling of protons \cite{gordon2014}.

To shed more light on the nature of noise and decoherence in Al$_2$O$_3$-based devices we use on-chip  electron spin resonance (ESR) spectroscopy in combination with various surface treatments. ESR is a non-destructive technique used to  probe the nature and concentration of paramagnetic centers and their interaction with the environment. In conventional ESR a sample is placed in a 3D cavity and the transitions between the spin states of the sample are detected by measuring the microwave energy absorption as a function of magnetic field. 
The figure of merit for conventional spectrometers is the total detectable number of spins in the sample, typically on the order of $10^{13}$, in a mm sized volume. Here we  interrogate spins using a recently developed and highly sensitive on-chip ESR technique \cite{aplpaper} based on magnetic field resilient planar superconducting NbN resonators fabricated on top of Al$_2$O$_3$. These on-chip spectrometers achieve better sensitivity owing to much higher quality (Q) factors and stronger coupling to the spins. 
For these on-chip spectrometers the relevant figure of merit is the minimum density of spins. We can resolve about $10^{13}$ spins per m$^2$.   
This enables us to probe the chemical nature of very dilute surface spins.

 \begin{figure*}
 	\includegraphics[scale=1.2]{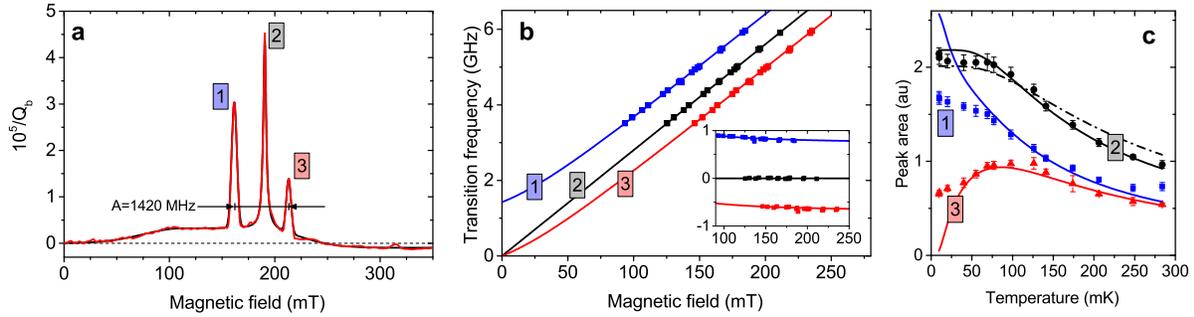}	 	
 	\caption{Spin energy spectrum and characteristics. a) Spectrum observed (red) at 10 mK with a fit (black). An excellent fit is obtained for a Lorentzian central peak and Gaussian satellite peaks. The broad spin background has a wide Gaussian onset followed by a constant contribution to the dissipation over almost 100 mT. The shown quantity $Q_b^{-1}=Q^{-1}(B)-Q^{-1}(B=0)$ is the magnetic field associated losses, obtained from the quality factor $Q$. b) Extracted peak positions for a large number of resonators with different frequencies jointly fitted to the energy level transitions of our model. Inset shows the same data where a constant slope of $g=2.0$ has been subtracted. c) Temperature dependence of the extracted peak areas fitted to the expected temperature dependence. Solid black line is for a doublet ($S=1/2$) spin ensemble while the black dashed line corresponds to the best fit to a triplet ($S=1$). Error bars are 95\% confidence bounds to spectral fits such as the fit presented in a).} \label{fig1}
 \end{figure*}

Fig. 1a shows the central result of this work: the ESR signal due to paramagnetic species on the annealed surface of  $\alpha-$Al$_2$O$_3$(0001). The measured ESR (red) and fit (black) show a Lorentzian central peak (line width $\gamma_2/2\pi=87$ MHz) accompanied  by two Gaussian satellites (line width of $\Delta/2\pi=90$ MHz) on top of a very wide spin signal 'pedestal', here with an onset at about 50 mT.

By measuring many devices and extracting the peak positions we are able to reconstruct the spin energy spectrum, shown in Fig. \ref{fig1}b. The simplest interpretation of the spectrum is that of two independent spectra: an ensemble of $S=1/2$ electron spins coupled to  $I=1/2$ nuclear spins together with a  free electron spin ensemble, justified by the excellent fit as well as the observed temperature dependence (see below).
The extracted hyperfine splitting is $A=1423.0\pm4.4$ MHz, precisely that of atomic hydrogen. Gyromagnetic ratios for both spin systems are $g=2.0$, with an absolute accuracy limited by screening and flux focussing effects in the superconducting resonator. The g-factors are, however,  the same to within at least 0.1\%.
  Rotating the magnetic field in the surface plane reveals  unchanged spin system parameters (Supplementary information), an indication that the spins are decoupled from the bulk Al$_2$O$_3$ crystalline field, which is also supported by the unperturbed hyperfine splitting of H. Remarkably, the hydrogen is weakly physisorbed on the surface. 

Further insight can be gained from the temperature dependence of the ESR transitions, shown in Fig. \ref{fig1}c. We obtain an excellent fit to the expected spin polarisation for our proposed spin system down to 50 mK. At zero temperature spins will only populate the ground state of their respective spin community. The existence of two non-vanishing peaks in the spectrum down to mK temperatures thus shows that there are two independent spin communities. Furthermore, the fit allows us to attribute the central peak to a $S=1/2$ doublet state. Other configurations ($S>1/2$) will result in a much weaker temperature dependence (see dashed line in Fig. \ref{fig1}c for a fit to $S=1$), this means that we can rule out for example triplet states in O$_2$ \cite{kumar2016} or atomic O \cite{harvey}, and the excellent fit to a Lorentzian line shape makes it unlikely that intrinsic nuclear spins, of for example Nb, N or Al \cite{wu2012}, are involved.

\begin{figure}[h!]
	\begin{center}
			\includegraphics[scale=1.1]{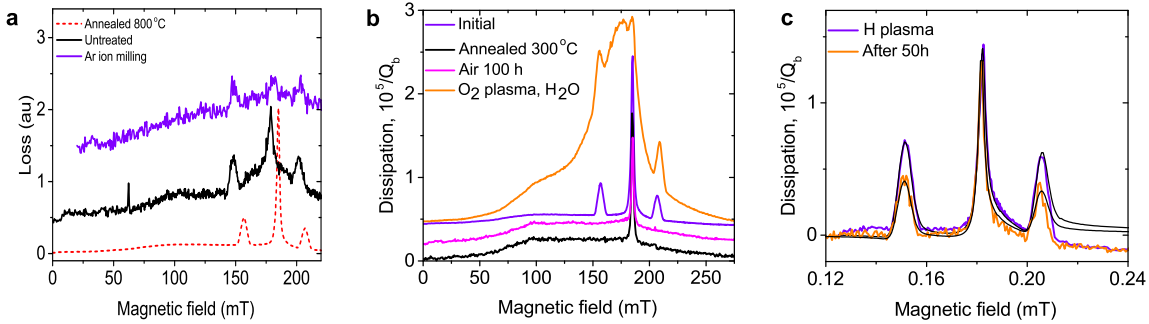}
	\end{center}
	\caption{Probing the surface chemistry.  a) The influence of a series of surface treatments (see text) on the measured spectrum. Curves have been offset for clarity. b) Post fabrication annealing a sample to 300$^\circ$ C removes the peaks associated with H via H$_2$ generation. The H does not return at ambient conditions. An Oxygen plasma and immersion into water returns the H-peaks with the same intensity, while the central peak remains largely unaffected throughout, even in the presence of the additional broad peak after immersion in water (Supplementary information). Traces have been offset for clarity. c) The result of exposing a sample to a H-ion plasma and subsequent relaxation back to the original spin density. The central peak is unaffected while the hydrogen spin density increased just after plasma exposure by $ 70$ \%, returning to equilibrium conditions after 50 hours. Solid black lines are fits to extract spin density (Supplementary information). $T = 300$ mK.} \label{surfacetreatment}
\end{figure}

In the sample presented in Fig. \ref{fig1}a the bare substrate was annealed at 800$^\circ$C prior to the deposition of NbN to improve the surface quality. 
Without this step the spectrum is significantly broader (Fig. \ref{surfacetreatment}a), and Ar ion milling, which further increases the surface roughness, results in an even broader spin resonance and an increased number of spins in both communities. 
It is thus clear that the spins are located on the surface of Al$_2$O$_3$, and are highly sensitive to the surface quality. 

From a detailed analysis of the number of spins interacting with our resonator (Supplementary information) we find a spin density of $n_H=1.2\pm0.5\times 10^{17}$ m$^{-2}$ of hydrogen, and $n_e=1.0\pm0.5\times 10^{17}$ m$^{-2}$ of free electron spins.
Remarkably, the total spin density is very close to the density thought to be responsible for flux noise in SQUIDs \cite{sendelbach2008,kumar2016}.
The observed abundance of the two species gives a ratio $n_H/n_e\approx 1.2$ (verified through the temperature dependence and through the collective coupling of the spins, Supplementary information). 

Finally, we have explored several treatment methods in order to affect the two spin communities.
Fig. 2b shows the initial spectrum and the spectrum after annealing the same sample at 300$^\circ$C. This  removes the hydrogen spins through an exothermic 2H$_{(ads)}\rightarrow$ H$_{2 (gas)}$ reaction, consistent with expected reaction barriers \cite{lee2014,lu2015}, and has a negligible effect on the free electron spins. 
 Long term exposure to ambient conditions with high humidity does not affect this state. Fig. 2b also shows the effect of exposing the sample to an oxygen plasma. After this step, exposure to water results in H-peaks reappearing with the same intensity as before. The additional broad resonance has a g-factor of $2.10\pm0.01$, and is most likely originating from excess H$_2$O adsorbed on the surface \cite{hass1998}.

Our measurements show that physisorbed atomic hydrogen is present on the surface, and to understand its origin we turn to the chemistry of the alumina surface.
 It is well known that dissociative H$_2$O adsorption on pristine Al-terminated $\alpha$-Al$_2$O$_3$(0001) leads to surface hydroxylation through an exothermic process \cite{lu2015,hass1998}. The adsorbed H$_2$O is split into OH$^-$ on the Al-site and H$^+$ is transferred to a nearby surface O. 
 Hydroxylation can occur through several different reaction pathways \cite{lu2015,hass1998,ma2016} and, importantly, pathways exist that lead to imperfections in the OH-terminated surface with several types of by-products, including unpaired H\cite{lu2015}. 
The perfect hydroxylated state has $\sim$20 OH/nm$^{2}$ and is expected to be ESR silent\cite{lee2014}. Such OH-terminated surface has been experimentally verified using several different techniques \cite{sung2011, eng2000}. Our observed surface spin density is a small fraction (1\%) of the total OH density and instead links to a stable density of physisorbed H, a by-product as the result of the hydroxylation process. From Fig. 2b it is evident that moderate annealing removes these atomic H. Since no further hydroxylation can occur on the fully OH-terminated surface no new H are  generated at ambient conditions.  Starting over (removing a large fraction of OH by an O-plasma), and exposing the Al$_2$O$_3$ surface to water results in the recovery of the same number of H. 

To verify that the amount of physisorbed H reduces to a stable concentration we exposed a sample to a H plasma (Fig. \ref{surfacetreatment}c). This led to an increase in the ratio $n_H/n_e= 2.0$ (without affecting the central peak), which after 50 hours exposure to atmosphere at 300 K was again reduced to $n_H/n_e= 1.3$, very close to the original value of 1.2. This clearly indicates that only excess physisorbed H leaves the surface and the density is not solely governed by a thermodynamic equilibrium but by some other mechanism. 

The origin of the central peak is more ambiguous, since $g=2.0$ is a characteristic of many different spin systems. Our findings still apply constraints on the nature of these spins. Molecular oxygen adsorbed on the Al-site was suggested as a plausible candidate for flux noise\cite{ kumar2016}. While the results in Ref. \cite{ kumar2016} do suggest that neutral oxygen is involved, the signature of a triplet state would have been clearly detected in our measurements. Our observation of a doublet state instead suggests that if oxygen is present it would be in charged molecular form, O$_2^-$ \cite{losee}.
Indicatively, the observed concentration of free electron  spins is very close to the density of Lewis base sites on Al$_2$O$_3$\cite{medvedev2012}. Although the exact nature of these defect-like sites (D$_s$) is still unknown, their presence is revealed by electron-donor reactions with different types of adsorbates and spin markers\cite{bedilo2014, medvedev2012}. We therefore argue that O$_2$ adsorbed via the reduction reaction O$_2+$D$_s^-\rightarrow$O$_2^-+$D$_s$ would explain the central peak. These spins are thus always present in our experiments since we could not avoid exposing the sample to air while mounting it in the cryostat.  However, we note that our data for the central peak itself does only indicate a single $S=1/2$ spin system with a g-factor of 2.0 on the surface, which would in principle be fulfilled for a large number of different adsorbates. 

We find it highly intriguing that the spin density for the two independent communities are stable at the same densities. It is possible that the concentration of H could be governed by the same $D_s$ sites, however, any additional insight into this mechanism so far remains elusive, prompting further experimental and theoretical investigation. 
Finally we also note that the broad spin signal with Gaussian onset and a flat wide plateau is present in all our experiments. The appearance of such signal may be explained by spin-spin interactions and clustering, a mechanism that have been considered as a likely candidate for 1/f flux noise \cite{schon2014}.

Our findings provide a straightforward path towards understanding and mitigating sources of noise in superconducting quantum circuits.
It has theoretically been found that hyperfine interactions provide a direct link to flux noise \cite{wu2012} and recent experimental results \cite{martinisfluxnoise2016} shows a clear peak in flux noise centred around 1.42 GHz, striking evidence that H produces noise even at zero applied magnetic field in superconducting qubits.
Furthermore, these results shed light on previous  studies showing an increase in number of TLS after exposing dielectrics \cite{jameson2011} and resonators \cite{khalil2013} to H-rich conditions, and saturation of Lewis donor sites with ESR-silent molecules could  explain the observed flux noise reduction in Ref. \cite{ kumar2016} which clearly provides a recipe for flux noise mitigation. 
The presented insight into the nature of two-level defects on Al$_2$O$_3$ may lead to new processing steps that will ultimately remove these defects and increase the coherence times of quantum circuits. The reported method may also be a valuable tool to better understand the surface chemistry of Al$_2$O$_3$ with implications for a wide range of applications.

\subsection*{Methods}
{\footnotesize
The samples used in this study were fabricated on C and R-cut sapphire substrates from different manufacturers. The fabrication technology is similar to that used in most solid state quantum device applications.

Before the deposition of 140 nm NbN samples were annealed in situ at high temperature, $800^\circ$C, for  20 minutes prior to deposition of 2 nm NbN. After cooling down to $20^\circ$C, an additional 140 nm NbN was sputtered. 
Magnetic field resilient resonators\cite{aplpaper} were patterned using electron beam lithography (UV60 resist, MF-CD-26 developer, DI water rinse) and subsequent reactive ion etching in a NF$_3$ plasma. Resist was removed in 1165 remover.  
The 'untreated', and 'Ar ion milling' samples in Fig. 1b were without the initial annealing step. The 'Ar ion milling' sample was exposed to an Ar ion milling step (300V, 5mA beam current, 3 minutes). All other ESR spectra were taken on high temperature annealed samples.
Surface treatments in Fig. 2 were carried out on samples annealed at $800^\circ$C. After cleaning the sample in 1165 remover and isopropanol for 12 hours it was annealed at $300^\circ$C for 15 minutes in high vacuum. It was exposed to ambient conditions for 2 hours while mounting in the cryostat. Additional exposure to ambient conditions after measurements for 50 hours had no effect on the spectrum. The same sample was then exposed to an oxygen plasma (250 W, 500 mTorr, 3 minutes) and soaked in DI water for 5 minutes.
One sample annealed to $800^\circ$C was subjected to a H-plasma (30 mBar H$_2$ pressure, ignition at 100 W, reduced to 10 W for 10 minutes) which affected only the two hydrogen peaks in the spectrum. Approximately 6 hours elapsed between plasma treatment and the first cooldown, during which the sample was exposed to ambient conditions. The same sample was then exposed another 50 hours to ambient conditions before being measured again.

Measurements were performed using microwave transmission spectroscopy in magnetic fields at milliKelvin temperatures in two different labs. All measurements were carried out using a vector network analyser and heavily attenuated microwave lines in the cryostat together with a cryogenic high electron mobility transistor amplifier with a noise temperature of $\sim 4$ K. The measured $S_{21}$ resonance data was fitted taking into account a complex coupling Q-factor using standard methods in order to extract the internal and external Q-factors, from which the magnetic field induced losses $Q_b$ were calculated for each applied magnetic field.
 Measurements were carried out at microwave probing powers well below spin saturation (see supplementary). Samples were mounted such that the applied static magnetic field was in plane with the superconducting film and orthogonal to the microwave magnetic field (see supplementary for a sketch).

}

\subsection*{Acknowledgements}
{\footnotesize
	The authors would like to thank B. Brennan for carrying out the SIMS analysis, S. Lara Avila for assistance with sample preparation and V. Shumeiko, D. Golubev, J. Burnett, J. Martinis and R. McDermott for fruitful discussions. This work was supported by the UK government's Department for Business, Energy and Industrial Strategy.
}


\clearpage\newpage

\begin{center}
	\Large Direct identification of dilute surface spins on Al$_2$O$_3$: Origin of flux noise in quantum circuits\\\vspace{4mm}{\bf Supplementary information}\\\vspace{4mm}\normalsize
	S. E. de Graaf$^{1\dagger}$, A. A. Adamyan$^2$, T. Lindstr\"om$^1$, D. Erts$^3$, S. E. Kubatkin$^2$, A. Ya. Tzalenchuk$^{1,4}$, and A. V. Danilov$^2$\\\vspace{2mm}\footnotesize
	$^1$ National physical laboratory, Hampton Road, Teddington, TW11 0LW, UK\\
	$^2$ Department of Microtechnology and Nanoscience, Chalmers University of Technology, SE-412 96 G\"oteborg, Sweden\\
	$^3$ Institute of Chemical Physics, University of Lativa, LV 1586, Latvia\\
	$^4$ Royal Holloway, University of London, Egham, TW20 0EX, UK\\
	$^\dagger$ sdg@npl.co.uk
\end{center}

\setcounter{linenumber}{1}
\appendix
\renewcommand\thefigure{\thesection S\arabic{figure}}    
\setcounter{figure}{0} 
\renewcommand\theequation{S\arabic{equation}}
\setcounter{equation}{0} 

\begin{figure}[ht!]
	\begin{center}	
		\includegraphics[scale=1.2]{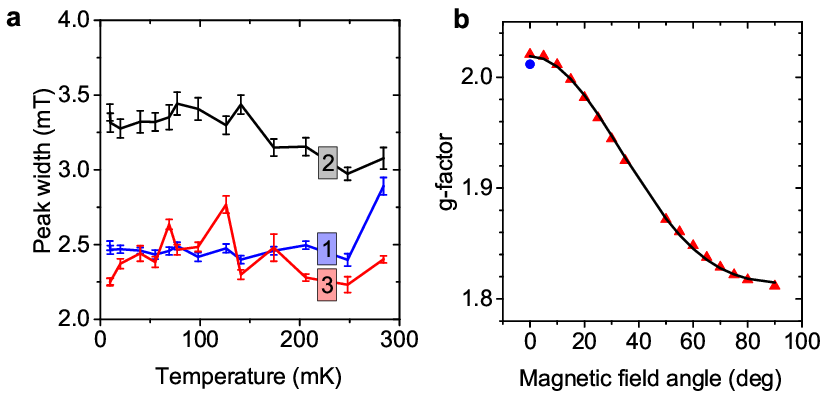}	
	\end{center}
	\caption{Details of the spectrum. a) The extracted peak linewidth (1 and 3: Gaussian, 2: Lorentzian) as a function of temperature. b) The measured g-factor as a function of the in plane magnetic field angle. Red datapoints are for a single sweep from 0 to 90$^\circ$, blue marker is obtained after returning from 90$^\circ$ and shows the effect of flux trapping and flux focussing. Solid line is a fit to $a\sin\theta+b\sin 2\theta$. All other spin ensemble parameters are  independent of angle and temperature in the measured range.} \label{esrlw}
\end{figure}

\begin{figure}[ht!]
	\begin{center}	
		\includegraphics[scale=0.55]{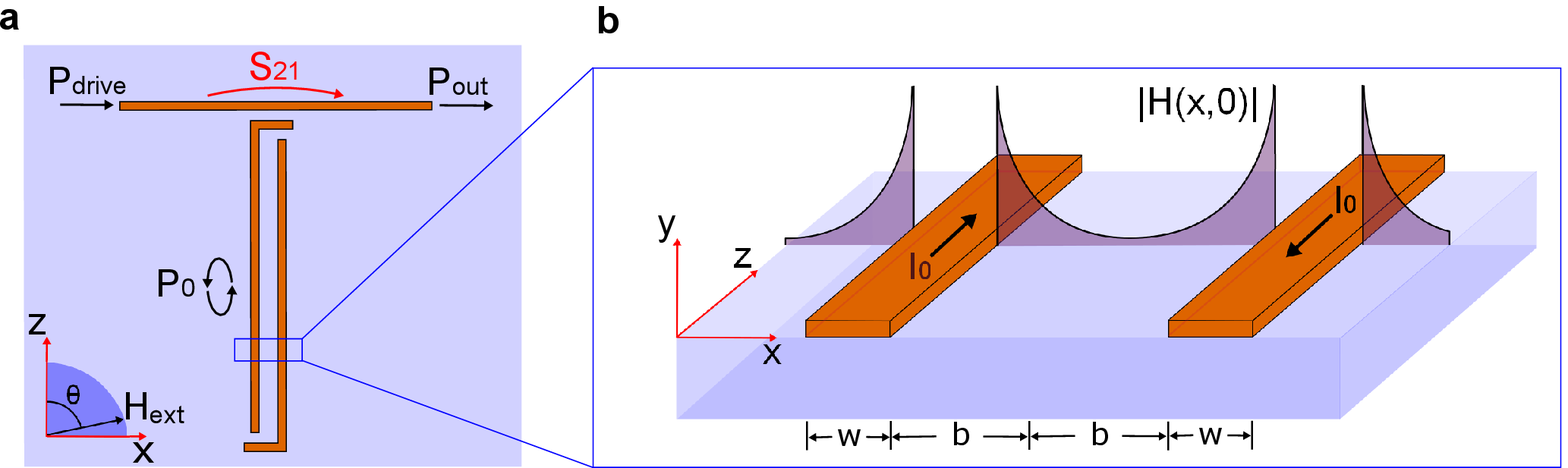}	
	\end{center}
	\caption{Geometry and working principle of the measurement. a) Simplified schematic of our ESR device (see \cite{aplpapersupp} for details). $H_{ext}$ shows the direction of the externally applied static magnetic field in the substrate plane, rotated with an angle $\theta$ with respect to the current carrying superconducting strips. An input power $P_{drive}$ results in a circulating power $P_0$ in the resonator. In the measurement the amplitude and phase of the transmission coefficient $S_{21}$ is monitored. b) The spin density is evaluated from the magnetic field $H$ in the substrate surface plane produced by the single photon current $I_0 = \sqrt{2\hbar\omega^2/Z}$ carried in the two parallel superconducting strips. } \label{sketchgeometry}
\end{figure}

\begin{figure}[ht!]
	\begin{center}		
		\includegraphics[scale=1.1]{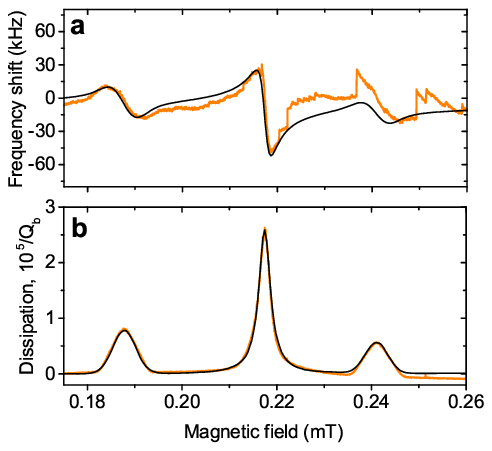}
		
	\end{center}
	\caption{Number of spins. Frequency shift (a) and dissipation (b) of a typical device and fit to eq. (S1) (black). $T=300$ mK. Sudden jumps in the measured frequency are related to flux avalanches in the superconducting ground plane surrounding the resonator.} \label{nofspins}
\end{figure}
\begin{figure}[ht!]
	\begin{center}	
		\includegraphics[scale=0.24]{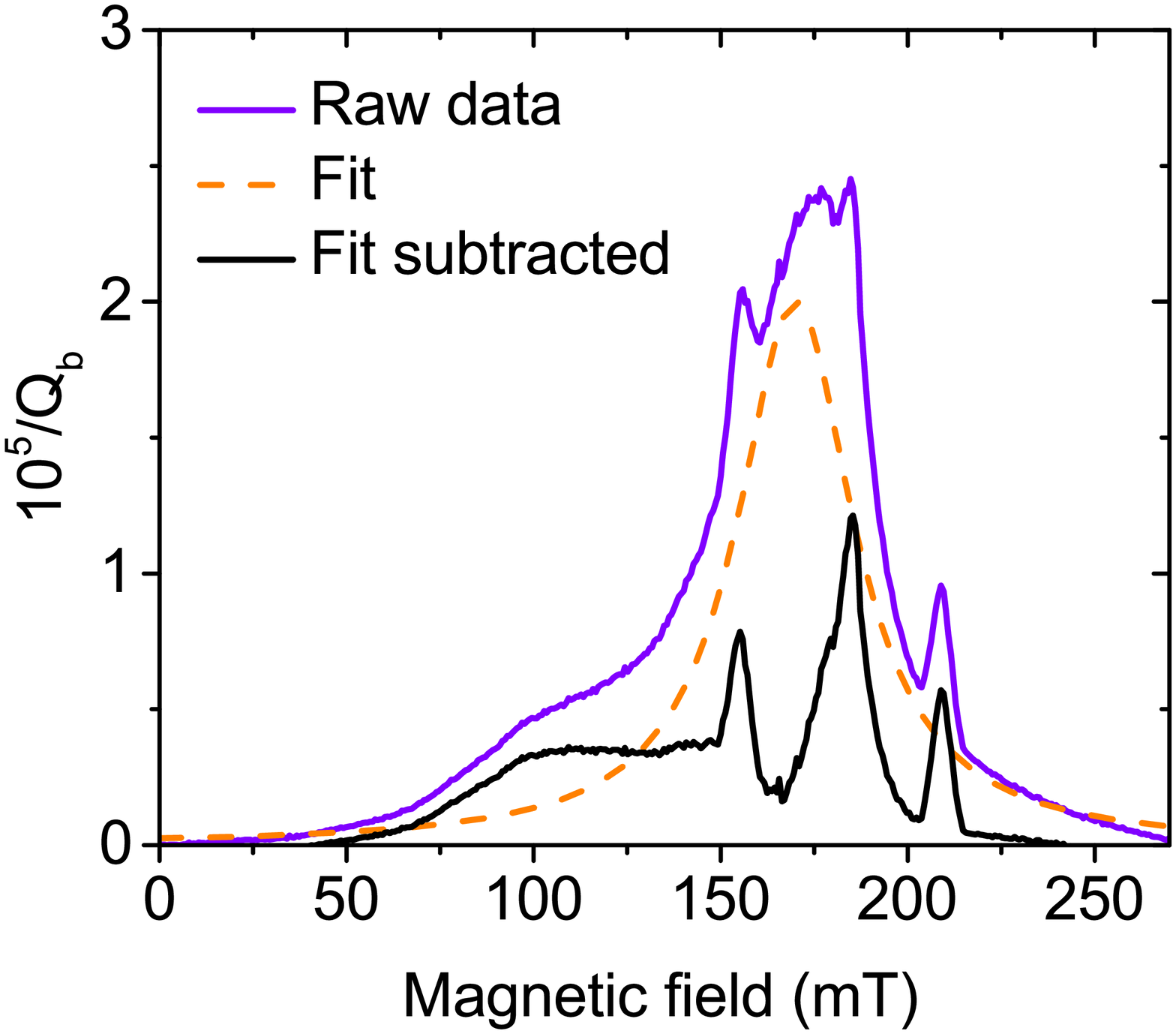}		
	\end{center}
	\caption{Spectral decomposition for the sample immersed in water. Subtracting the broad additional peak reveals the three original peaks of similar intensities.} \label{h2odecomposed}
\end{figure}

\begin{figure}[h!]
	\begin{center}	
		\includegraphics[scale=0.27]{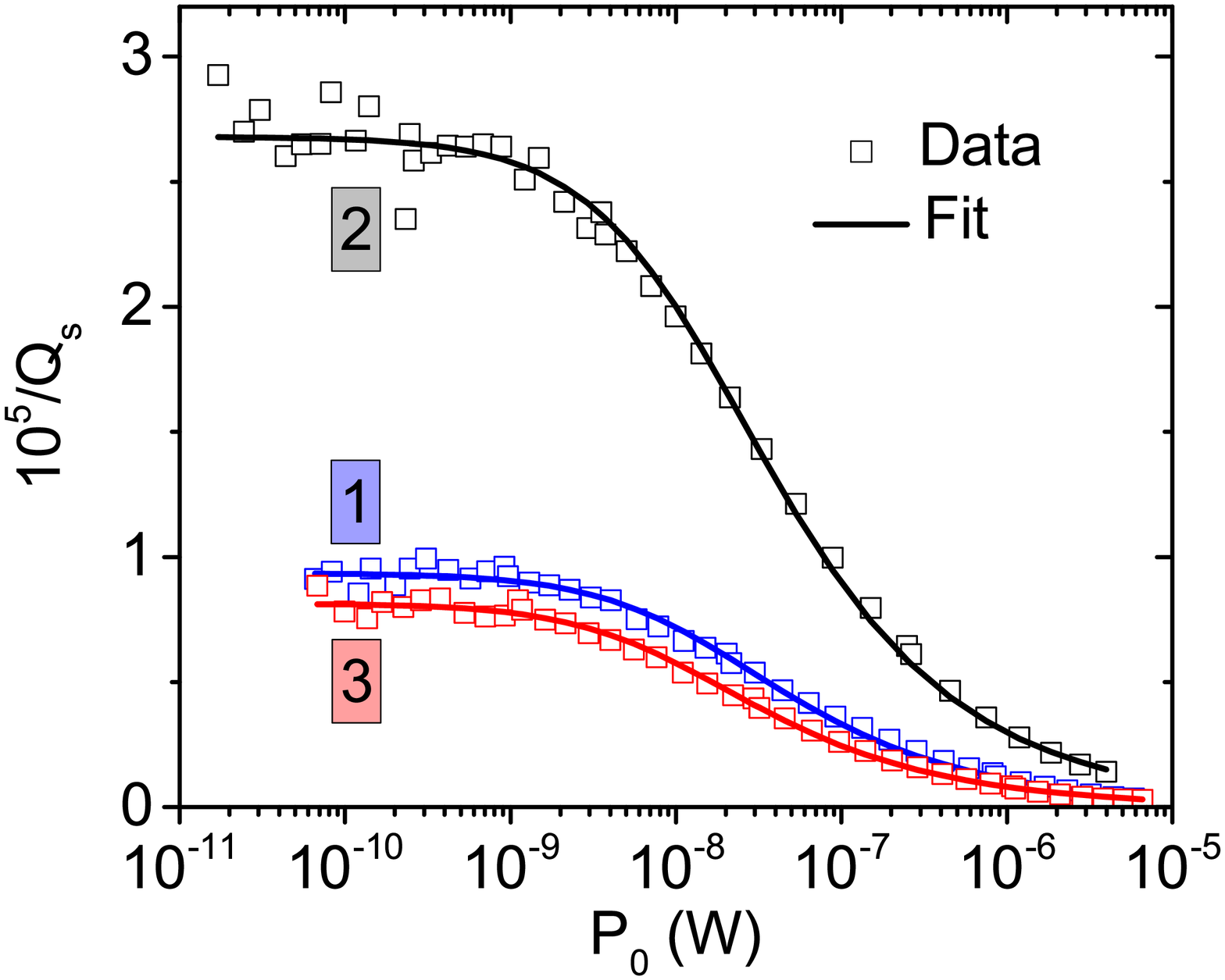}		
	\end{center}
	\caption{CW power saturation. Quality factor associated with the dissipation of energy into the spin system as a function of circulating power in the resonator at $T=300 $ mK.} \label{nofspinssat}
\end{figure}

\begin{figure}[h!]
	\begin{center}	
		\includegraphics[scale=0.33]{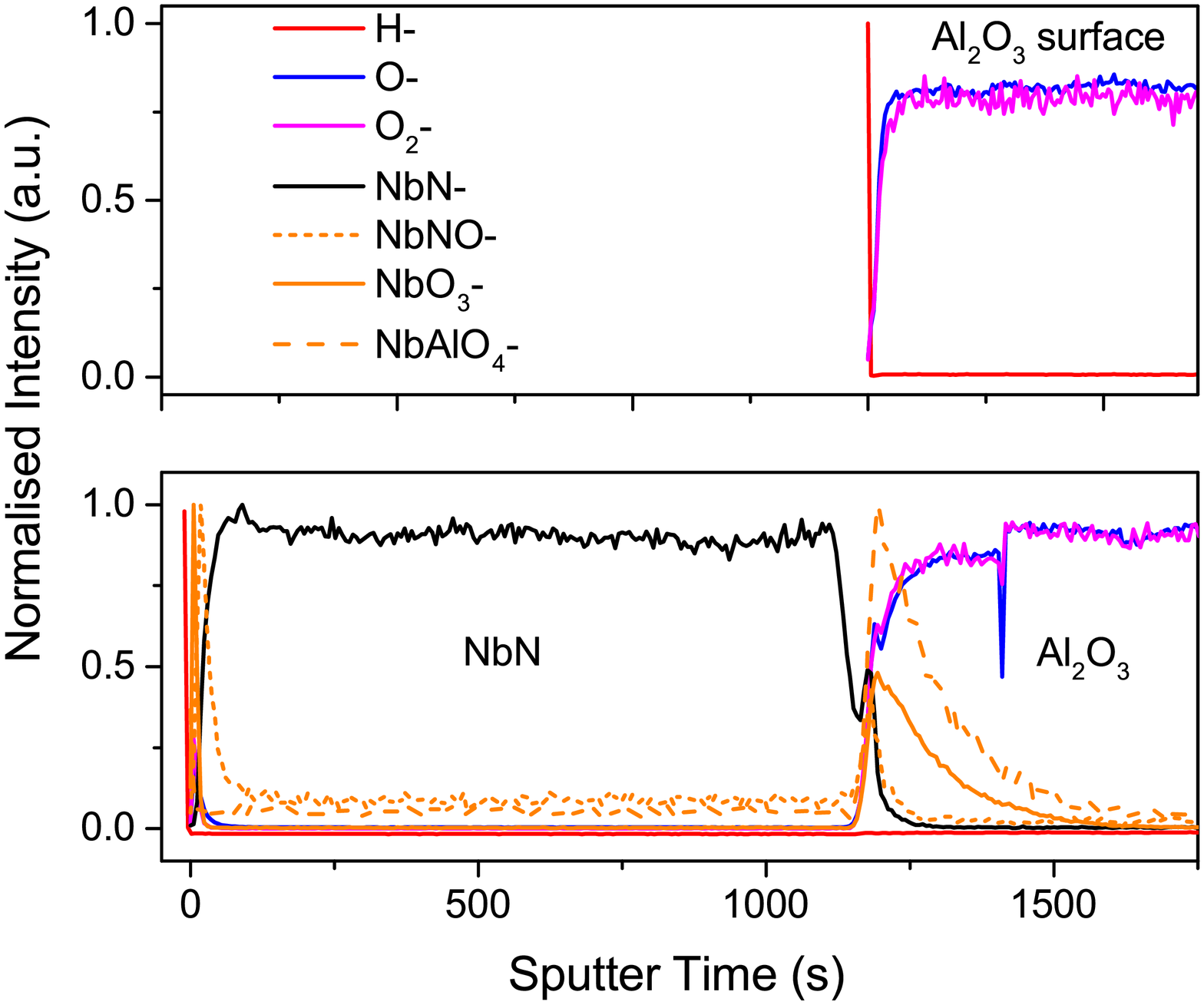}		
	\end{center}
	\caption{Secondary Ion Mass Spectrometry (SIMS) analysis. Top: The bare Al$_2$O$_3$ surface. Bottom: A region covered by NbN. } \label{sims}
\end{figure}

\clearpage

\subsection*{Other properties of the ESR spectrum}
We here present additional data concerning the properties of the measured spin systems. Fig. \ref{esrlw}a  shows the extracted line-width as a function of temperature. No temperature dependence was found for the hyperfine splitting energy, $A$, or the g-factor. Fig. \ref{esrlw}b shows the effect of rotating the applied static magnetic field in plane with the Al$_2$O$_3$ surface using a vector magnet. We attribute the observed changes to screening due to the superconductor. 
When the applied static magnetic field is parallel to the lines carrying the microwave current ($\theta=0$, see Fig. \ref{sketchgeometry}) in our resonator there is no demagnetization effect on the magnetic field applied to the spins. In contrast, while when the field is normal to the current carrying lines ($\theta = 90^\circ$) the static field is partially screened which gives an apparent shift in the g-factor. The true g-factor is thus the maximum obtained at $\theta =0$ in Fig. \ref{esrlw}b. In our highly field resilient resonators \cite{aplpapersupp} flux trapping is reduced to a minimum, and can be further reduced by sweeping the field back and forth many times prior to measuring the spectrum. However, an absolute uncertainty in the g-factor of $\sim 0.03$ due to variations in trapped flux still exists, as shown by the additional data point at $\theta =0$ in Fig. \ref{esrlw}b obtained after measuring all the other data points. 
No angular dependence was found for the hyperfine splitting or line-width of the peaks (data not shown). 

The nuclear spin saturation below 50 mK may be an indication of spin-spin interactions \cite{sendelbach2008supp} or quantum tunnelling  of protons \cite{holder2013supp, gordon2014supp}, and it is interesting to search for such signatures in the ESR spectrum.
In our present samples we are unable to detect any such signatures. This does not exclude but only gives an upper limit  of $\lesssim 50$ mK to the energy scales for such interactions.

\subsection*{Number of spins}
The measured frequency shift and change in quality factor of the resonator are fitted to the expected transmission
\begin{equation}
S_{21}(\omega) = 1+\frac{\kappa_c}{i(\omega-\omega_0)-\kappa+W_L(\omega)+W_G^\pm(\omega)},
\end{equation}
where the Lorentzian central peak is described by
\begin{equation}
W_L(\omega) = \frac{\Omega^2}{i(\omega-\omega_s)-\gamma_2/2},
\end{equation}
and the two Gaussian satellite peaks are given by
\begin{equation}
W_G^\pm(\omega) = \frac{\Omega^2\sqrt{\ln{2}}}{\Delta}\mathcal{W}\left(\frac{\omega-\omega_s^\pm+i\gamma_2/2}{\Delta/\sqrt{\ln{2}}}\right),
\end{equation}
where $\mathcal{W}(z)$ is the Faddeeva function\cite{diniz2011}. $\omega_0, \kappa=\omega_0/ Q, $ and $\kappa_c=\omega_0/ Q_c$ are the bare angular resonance frequency, total, and coupling decay rates respectively of the superconducting resonator.

At 300 mK, based on fits to eq. (S1), we get that the first satellite contains 22.5\% less spins than in the central peak and the right satellite has 57.6\% less spins. In total there are 19.9\% more H than free electron spins. From the temperature dependence we find the abundance $n_H/n_e= 1.195$. The inhomogeneous broadening for the satellite peaks is found to be $\Delta = 90$ MHz. For each peak we extract the collective coupling strength $\Omega$ and the pure dephasing rate $\gamma_2 = 2\pi/T_{2e}$. The surface density of spins is given by 
\begin{equation} n = \frac{8\Omega^2\pi^2\hbar^2}{\beta(T)\mu_B^2L_{res}}\left(\int_\delta |H(x,y=0)|^2dx\right)^{-1},\end{equation} 
where the integration is over a distance much longer than the extent of the electromagnetic field, $\beta(T) = (1-e^{\hbar\omega/k_BT})/(1+e^{\hbar\omega/k_BT})$ is the thermal spin polarization factor and

\begin{equation} H(x,y) = -\frac{\sqrt{2\hbar\omega^2/Z}}{2SK(\sqrt{1-b^2/S^2})}\frac{S^2}{\sqrt{((x+iy)^2-b^2)((x+iy)^2-S^2)}}\end{equation} 
is the magnetic field intensity produced by a single photon excitation in the resonator for our device geometry. $L_{res}$, $w$, and $2b$ are the total length of superconducting strip carrying current, its width, and separation to the second parallel strip carrying opposite current respectively and $S=b+w$, as sketched in Fig. \ref{sketchgeometry}. $K(z)$ is the elliptic function, and $Z$ the resonator impedance.
Example fits to eq. (S1) are shown in Fig. \ref{nofspins}a and b. 
Finally, Fig. \ref{h2odecomposed}  shows that the dissipation signal for all three peaks after water immersion are close to the original number  prior to re-hydroxylation.

\subsection*{CW power saturation}
In Fig. \ref{nofspinssat} we show the inverse quality factor due to the spin system as a function of circulating power in the resonator. This quantity is obtained from the difference in measurements of the power dissipated on and off resonance with the spin system, $P_{\rm{s}} = P_{\rm{on}}-P_{\rm{off}}$.

From the data in Fig. \ref{nofspinssat} it is clear that for a circulating power $P_0=2Q^2P_{drive}/Q_{ext}$ below 1 nW the spin system is not overheated by microwave excitation. This we use to verify that further analysis was done with the spin system at thermal equilibrium.

At higher excitation powers, when the excitation power is driving the spins out of equilibrium, the spin-related dissipation is expected
to have the following dependence on the circulating power \cite{haas1993supp}:
\begin{equation} 
\frac{1}{Q_{s}(P_0)} =\frac{P_{\rm{s}}}{P_0}= \frac{1}{Q_{s}(P_0=0)}\frac{1}{(1+P_0/P_{sat})^\epsilon}\end{equation} 
where $P_{sat} = 1/(T_{1e}T_{2e}\gamma_e^2\alpha^2)$, $\gamma_e$ is the gyromagnetic ratio and $\alpha = H/\sqrt{P_0}$ is a microwave power to microwave magnetic field conversion coefficient. 

From the fit we find $P_{sat} = (14,13,10)$ nW and $\epsilon =1$, for the three different peaks respectively. Due to magnetic field non-uniformity our superconducting resonator has a conversion factor that differs significantly from a conventional ESR spectrometer. We find an approximate value for two parallel superconducting strips
\begin{equation}\alpha = \frac{\mu_0}{2\sqrt{2Z} S K(\sqrt{1-b^2/S^2})}\frac{1}{w}\int_{S}^{S+w}\left|\frac{S^2}{\sqrt{(x^2-b^2)(x^2-S^2)}}\right|dx\approx 0.21 \hspace{2mm}\rm T/\sqrt{W},\end{equation} 
where we assume that most of the magnetic energy is localised to within one width of the superconducting strip. This value is about 3 orders of magnitude larger than in conventional ESR spectrometers, an indication of the strong coupling that can be obtained to individual spins and the high sensitivity of our technique.

The Lorentzian line shape allows us to directly relate the line width of the central peak to $T_{2e} = 12$ ns, and we can thus estimate $T_{1e}=200$ $\mu$s.

\subsection*{SIMS data}
To further characterise our samples we used Secondary Ion Mass Spectrometry (SIMS), where an ion beam (here 25 keV Bi$^{3+} $ 0.1 pA analysis beam, 10 keV Cs$^-$ 30 nA sputter beam) is used to locally sputter the material under study, and the sputtered species are analysed in a mass spectrometer vs sputtering time. Figure \ref{sims} shows the normalised relative intensity of selected species from SIMS measurements. Fig. \ref{sims}a shows data from a region of bare Al$_2$O$_3$ while Fig. \ref{sims}b shows a region of Al$_2$O$_3$ covered by NbN. 

\end{document}